\pdfoutput=1
\documentclass[11pt]{article}
\usepackage[]{acl}
\usepackage{times}
\usepackage{enumitem}
\usepackage{tikz}
\usetikzlibrary{shapes, arrows.meta, positioning}
\usepackage{booktabs}
\usepackage{multirow}
\usepackage{latexsym}
\usepackage{amsmath}  
\usepackage{amssymb}   
\usepackage{tikz} 
\usepackage{xcolor}
\usepackage[T1]{fontenc}
\usepackage[utf8]{inputenc}
\usepackage{microtype}
\usepackage{inconsolata}
\usepackage{graphicx}
\usepackage{ subcaption, algorithm2e, url, float, etoolbox, adjustbox, pgf,booktabs,verbatim}
\usepackage{xcolor,graphicx,colortbl}

\definecolor{litepurple5}{RGB}{237,231,246}   
\definecolor{litepurple10}{RGB}{209,196,233}
\definecolor{litepurple15}{RGB}{179,157,219}
\definecolor{litepurple20}{RGB}{149,117,205}
\definecolor{litepurple25}{RGB}{126,87,194}    

\title{\texttt{DIVINE} : Coordinating Multimodal Disentangled Representations for Oro-Facial Neurological Disorder Assessment}

\author{
 \textbf{Mohd Mujtaba Akhtar\textsuperscript{1}\thanks{Equal contribution as a first author.}},
 \textbf{Girish\textsuperscript{2}\footnotemark[1]} and
 \textbf{Muskaan Singh\textsuperscript{3}\thanks{Corresponding: \href{mailto:m.singh@ulster.uk.in}{m.singh@ulster.uk.in}}} \\
 \textsuperscript{1}Veer Bahadur Singh Purvanchal University, India \\
 \textsuperscript{2}UPES, India \\
 \textsuperscript{3}Ulster University, UK
}

\begin{document}
\maketitle

\begin{abstract}
In this study, we present a multimodal framework for predicting neuro-facial disorders by capturing both vocal and facial cues. We hypothesize that explicitly disentangling shared and modality-specific representations within multimodal foundation model embeddings can enhance clinical interpretability and generalization. To validate this hypothesis, we propose \textbf{DIVINE} a fully disentangled multimodal framework that operates on representations extracted from state-of-the-art (SOTA) audio and video foundation models, incorporating hierarchical variational bottlenecks, sparse gated fusion, and learnable symptom tokens. \textbf{DIVINE} operates in a multitask learning setup to jointly predict diagnostic categories (Healthy Control, ALS, Stroke) and severity levels (Mild, Moderate, Severe). The model is trained using synchronized audio and video inputs and evaluated on the Toronto NeuroFace dataset under full (audio-video) as well as single-modality (audio-only and video-only) test conditions. Our proposed approach, \textbf{DIVINE} achieves SOTA result, with the DeepSeek-VL2 and TRILLsson combination reaching \textsc{98.26}\% accuracy and \textsc{97.51}\% F1-score. Under modality-constrained scenarios, the framework performs well, showing strong generalization when tested with video-only or audio-only inputs. It consistently yields superior performance compared to unimodal models and baseline fusion techniques. To the best of our knowledge, \textbf{DIVINE} is the first framework that combines cross-modal disentanglement, adaptive fusion, and multitask learning to comprehensively assess neurological disorders using synchronized speech and facial video. 

\end{abstract}

\section{Introduction}
Neurodegenerative and neurovascular conditions such as Amyotrophic Lateral Sclerosis (ALS) and stroke often arise with impairments in facial motor control and speech articulation—symptoms that are not only diagnostic indicators but also indicative of disease progression \cite{bandini2020new,naeini2022automated}. 
\begin{figure}[!bt]
    \centering
    \includegraphics[width=0.9\linewidth]{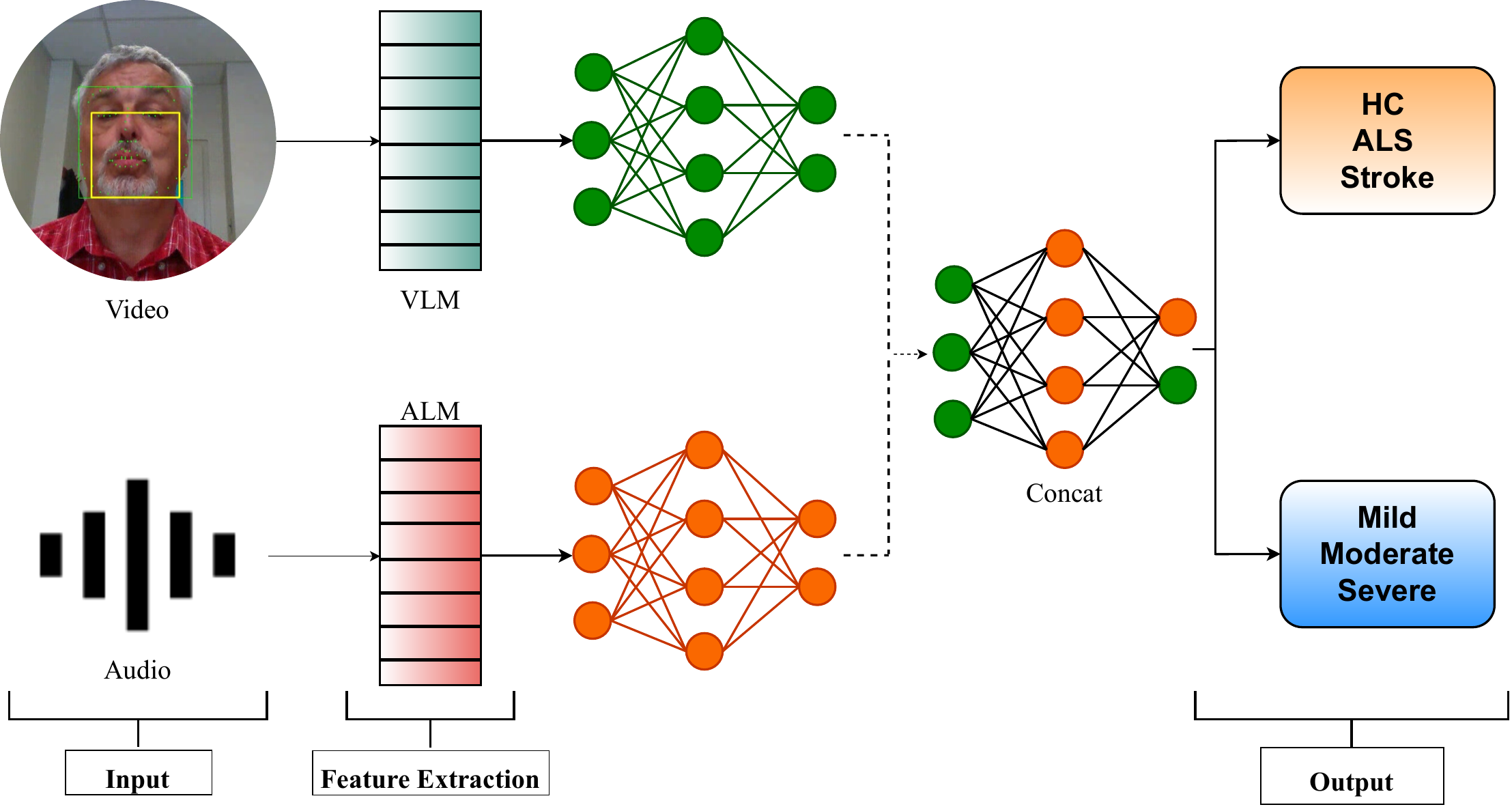}
    \caption{Overview of the study setup.}
    \label{fig:1fgd}
\end{figure}
Current clinical evaluations of these symptoms rely heavily on subjective expert assessments, which are labor-intensive, variable across raters, and difficult to scale for longitudinal monitoring. Recent computer vision and speech processing advances have demonstrated promising capabilities in analyzing facial kinematics and vocal patterns for clinical inference. In particular, leveraging facial landmarks \cite{gomes2023facial} and acoustic modeling \cite{migliorelli2023store} have enabled more objective quantification of motor dysfunction in neuro-facial disorders. However, these efforts often treat each modality in isolation, neglecting the complementary nature of audiovisual cues and their temporal co-dynamics in pathological speech and gestures. In contrast, multimodal architectures provide a more robust and holistic solution by \textit{jointly leveraging visual and acoustic information}. Nevertheless, earlier fusion strategies frequently struggle to separate modality-specific patterns from shared cross-modal representations. This limitation hampers both interpretability and generalizability, key requirements for ensuring clinical reliability. As illustrated in Figure~\ref{fig:1fgd}, our task uses synchronized speech and facial video to predict both the diagnostic category and the associated severity level. \par
To address the limitations of prior multimodal approaches, we propose \texttt{DIVINE} (\textbf{DI}sentangled \textbf{V}ariational \textbf{IN}formation \textbf{N}\textbf{E}twork), a fully disentangled, multitask audio-visual framework for the assessment of neuro-facial disorders. \texttt{DIVINE} integrates pretrained foundation models for both audio and video modalities and employs a hierarchical variational bottleneck to disentangle private (modality-specific) and shared (cross-modal) latent representations. It introduces a sparse gated fusion mechanism that dynamically modulates the influence of each modality and a symptom-guided tokenisation module that directs attention to clinically salient oro-motor features. 
\textit{We hypothesise that explicitly disentangling shared and modality-specific latent information enhances both disorder classification and severity estimation, while improving generalisation across diverse clinical tasks and input types}. To test this, we conduct extensive evaluations on three clinical populations—HC, ALS, and stroke survivors—across speech, non-speech, and mixed-task conditions. Our model performs multitask learning to jointly predict disorder type and five clinician-rated perceptual severity scores. Through systematic ablations and modality dropout experiments, we demonstrate that \textbf{DIVINE} maintains top performance under unimodal (audio-only, video-only) and multimodal conditions, establishing a new benchmark in multimodal neuro-facial assessment. \par
\noindent \textbf{To summarize, the main contributions of our study are:}
\begin{itemize}
    \item We introduced \textbf{DIVINE} (\textbf{DI}sentangled \textbf{V}ariational \textbf{IN}formation N\textbf{E}twork), a fully disentangled audio–visual variational framework that employs hierarchical variational bottlenecks, cross-modal alignment, gated fusion blocks, and symptom-token modules to extract and integrate complementary speech and facial representations for joint diagnosis and continuous severity estimation of neuro-facial disorders.
\vspace{-0.3mm}
    \item  We validate our framework on the Toronto NeuroFace dataset under three evaluation settings—full-modality (both audio and video inputs), partial-modality (speech-only or non-speech-only segments), and missing-modality (audio-only or video-only inputs)—and also benchmark over 40 combinations of SOTA audio and vision foundation models.

    \item To the best of our knowledge, \textbf{DIVINE} is the first unified framework to combine hierarchical disentangled latent learning, cross-modal alignment losses, and multitask objectives—simultaneously addressing categorical classification (Healthy Control, ALS, Stroke) and regression-style severity prediction—in a single, end-to-end pipeline.
\end{itemize}

\noindent \texttt{Project resources (code, model checkpoints, and evaluation scripts) are available at:\footnote{\url{https://github.com/Helixometry/SIGNAL.git}}}

\section{Related Work}
\vspace{-0.3cm}
Early work in oro-facial neurological assessment relied solely on video or images. Researchers used handcrafted spatio-temporal features, such as Improved Dense Trajectories with Fisher Vector encoding, to capture broad facial movements in natural settings \cite{wang2013action,afshar2016facial}. \cite{bandini2020new} introduced the NeuroFace benchmark, showing that standard face-alignment tools can struggle with pathological motion. More recent methods apply deep models: maximisation–differentiation networks for depression screening \cite{de2021mdn}, multiscale CNNs for expression analysis \cite{de2024facial}, and landmark-aware transformers for estimating expression intensity \cite{chen2024static}. Graph neural networks have also been used to model facial asymmetry and rigidity in ALS patients by treating landmarks as nodes in a facial graph \cite{gomes2023facial}. To address video’s limitations (occlusion, lighting), simple fusion approaches combine visual and acoustic cues. Related audio-visual speech recognition work has also shown that explicitly conditioning strong audio models on visual features via attention-based fusion can improve robustness under noisy acoustics \cite{rouditchenko24_interspeech,choi-etal-2025-leveraging}.
\cite{duan2023qafe} proposes a two-stream system that fuses landmark heat-map volumes with RGB frames via a cross-fusion decoder, improving motion capture. \cite{neumann2024multimodal} builds a remote dialog system that extracts facial, linguistic, and acoustic biomarkers from ALS patients to track bulbar decline over time. While these methods combine modalities, they treat all features as a single block without separating what each modality contributes. More recent research aims to learn separate, meaningful factors and tackle multiple tasks simultaneously \cite{duan2023qafe, neumann2024multimodal}. \cite{shi2019variational} further explores Variational Mixture‐of‐Experts Autoencoder (MMVAE), which factorises the joint posterior as a mixture of unimodal experts to disentangle shared and private latents (modality-specific) and support coherent multi‐modal inference. Our work departs from these by introducing a fully disentangled multimodal framework that separates private (audio‐ or video‐specific) and shared representations, and supports joint diagnosis and severity estimation. This approach allows us to quantify each modality’s contribution and handle missing or noisy inputs more robustly than previous fusion strategies.

\section{Representations}
\noindent\textbf{Speech Models} Our speech encoders include monolingual models—\textit{Wav2Vec2.0}~\cite{baevski2020wav2vec} and \textit{WavLM}~\cite{chen2022wavlm}—trained on large-scale English corpora using contrastive and masked prediction objectives. We also leverage \textit{HuBERT}~\cite{hsu2021hubert}, which predicts latent acoustic units via masked prediction, capturing long-range dependencies in speech.
We also include multilingual models such as \textit{Whisper}~\cite{radford2023robust}, trained on 680k hours of cross-lingual data, trained on 128 languages. For prosodic variation and speaker-dependent cues, we use \textit{TRILLsson}~\cite{shor2022trillsson} and \textit{x-vector}~\cite{8461375}, both known for their robustness in paralinguistic speech tasks. \newline
\noindent\textbf{Vision Models} For facial video modeling, we utilize transformer-based models including \textit{VideoMAE}~\cite{tong2022videomae}, \textit{VideoMAE-V2}~\cite{wang2023videomae}, and \textit{ViViT}~\cite{arnab2021vivit}, all employing spatiotemporal encoding strategies. We further use \textit{DeepSeek-VL2}~\cite{wu2024deepseek}, a vision-language model with a dynamic tiling and token aggregation mechanism. As structured baselines, we include handcrafted kinematic features from OpenFace~\cite{8373812} and temporal attention features extracted using a ResNet18+TANN pipeline.
\noindent Additional details regarding the above pretrained models (PTMs) are provided in Appendix~\ref{PTMS}.
\begin{table*}[!t]
\setlength{\tabcolsep}{8pt}
\scriptsize
\centering
\begin{tabular}{l|cc|cc|cc|cc|cc|cc}
\toprule
\multicolumn{1}{l}{} 
  & \multicolumn{2}{c}{\textbf{FCN}} 
  & \multicolumn{2}{c}{\textbf{CNN}} 
  & \multicolumn{2}{c}{\textbf{FCN}} 
  & \multicolumn{2}{c}{\textbf{CNN}} 
  & \multicolumn{4}{c}{\textbf{CNN}} \\
\cmidrule(lr){2-3} \cmidrule(lr){4-5} \cmidrule(lr){6-7} \cmidrule(lr){8-9} \cmidrule(lr){10-13}
\multicolumn{1}{l}{} 
  & \textbf{A \(\uparrow\)} & \textbf{F1 \(\uparrow\)} 
  & \textbf{A \(\uparrow\)} & \textbf{F1 \(\uparrow\)} 
  & \textbf{M \(\downarrow\)} & \textbf{R \(\downarrow\)} 
  & \textbf{M \(\downarrow\)} & \textbf{R \(\downarrow\)} 
  & \textbf{A \(\uparrow\)} & \textbf{F1 \(\uparrow\)} & \textbf{M \(\downarrow\)} & \textbf{R \(\downarrow\)} \\
\midrule
\multicolumn{1}{l}{} 
  & \multicolumn{4}{c}{\textbf{C}} 
  & \multicolumn{4}{c}{\textbf{R}} 
  & \multicolumn{4}{c}{\textbf{M}} \\
\cmidrule(lr){2-5} \cmidrule(lr){6-9} \cmidrule(lr){10-13}
\multicolumn{13}{c}{\textbf{VFM}} \\
\midrule
\textbf{Vi}  & 80.54 & 78.57 & 83.78 & 81.35 & 10.82 &  8.99 &  9.76 &  7.28 & 82.89 & 82.62 & 10.25 &  8.64 \\
\textbf{V2}  & \underline{82.16} & \underline{81.65} & \underline{85.69} & \underline{83.58} & \underline{ 9.29} & \underline{ 7.38} & \underline{ 8.73} & \underline{ 6.84} & \underline{85.38} & \underline{84.81} & \underline{ 9.45} & \underline{ 7.65} \\
\textbf{VV}  & 79.16 & 77.28 & 82.59 & 81.27 & 11.26 &  9.52 & 10.22 &  8.63 & 81.53 & 80.16 & 11.86 &  9.32 \\
\textbf{DS}  & \textbf{85.23} & \textbf{84.61} & \textbf{88.94} & \textbf{86.57} & \textbf{ 9.22} & \textbf{ 7.26} & \textbf{ 8.58} & \textbf{ 6.81} & \textbf{88.33} & \textbf{86.15} & \textbf{ 9.38} & \textbf{ 7.58} \\
\textbf{KI}  & 72.29 & 71.61 & 76.16 & 74.51 & 11.82 &  9.58 & 10.50 &  8.89 & 75.45 & 73.98 & 11.55 &  9.88 \\
\textbf{TA}  & 78.31 & 77.18 & 79.56 & 77.06 & 10.58 &  8.97 &  9.96 &  8.05 & 78.11 & 76.65 & 11.09 &  9.11 \\
\midrule
\multicolumn{13}{c}{\textbf{SFM}} \\
\midrule
\textbf{WV}  & 78.29 & 77.19 & 80.83 & 79.37 &  8.38 &  9.70 &  7.61 &  8.51 & 82.09 & 80.35 &  6.88 &  7.60 \\
\textbf{W2}  & 74.02 & 73.77 & 76.37 & 74.32 &  8.54 &  9.89 &  7.66 &  8.38 & 82.13 & 81.98 &  6.38 &  7.55 \\
\textbf{WR}  & 80.61 & 79.49 & 82.34 & 81.06 &  8.47 &  9.36 &  7.18 &  8.16 & 85.94 & 83.56 &  6.22 &  7.29 \\
\textbf{XV}  & \underline{85.85} & \underline{83.96} & \underline{86.29} & \underline{85.81} & \underline{ 8.16} & \underline{ 8.59} & \underline{ 6.94} & \underline{ 7.61} & \underline{89.27} & \underline{87.64} & \underline{ 6.15} & \underline{ 7.14} \\
\textbf{HT}  & 77.39 & 76.28 & 79.62 & 78.09 &  9.72 & 10.12 &  8.68 &  9.87 & 80.11 & 79.51 &  6.85 &  7.74 \\
\textbf{TR}  & \textbf{86.06} & \textbf{84.64} & \textbf{87.58} & \textbf{86.64} & \textbf{ 7.50} & \textbf{ 7.88} & \textbf{ 6.83} & \textbf{ 7.25} & \textbf{90.51} & \textbf{88.69} & \textbf{ 6.12} & \textbf{ 7.01} \\
\bottomrule
\end{tabular}
\caption{Performance of individual Video Foundation Models (VFMs) and Speech Foundation Models (SFMs) on speech-video samples across classification, regression, and multitask settings using FCN and CNN heads. \textbf{Best} and \underline{second-best} results in each column are highlighted (same convention in Table~\ref{tab:2}. \textbf{Abbreviations:} VFMs—Vi (VideoMAE), V2 (VideoMAE V2), VV (ViViT), DS (DeepSeek-VL2), KI (Kinematic), TA (Temporal); SFMs—WV (WavLM), W2 (Wav2Vec2), WR (Whisper), XV (X-vector), HT (HuBERT), TR (TRILLsson). {\textcolor{blue}{Note: The abbreviations used in Table~\ref{tab-1} are consistent across Tables~\ref{tab:2}, \ref{tab:3},\ref{tab:4},\ref{tab:5} and \ref{tab:6}.}}}
\label{tab-1}
\end{table*}

\section{Modeling}
We consider two downstream networks, i.e., a fully connected network (FCN) and a CNN with individual PTM representations applied independently to each audio and video foundation model representation. The FCN model consists of three dense layers with 256, 128, and 64 neurons, followed by the output layer. The CNN model comprises two convolution blocks, each containing a 1D convolutional layer followed by batch normalization and a max-pooling operation, then a flattening step and a dense FCN block with the same configuration as above.  Detailed hyperparameter settings and model configurations are described in Appendix \ref{hyperparameter}.

\begin{figure*}[!hbt]
    \centering
    \includegraphics[width=0.9\linewidth]{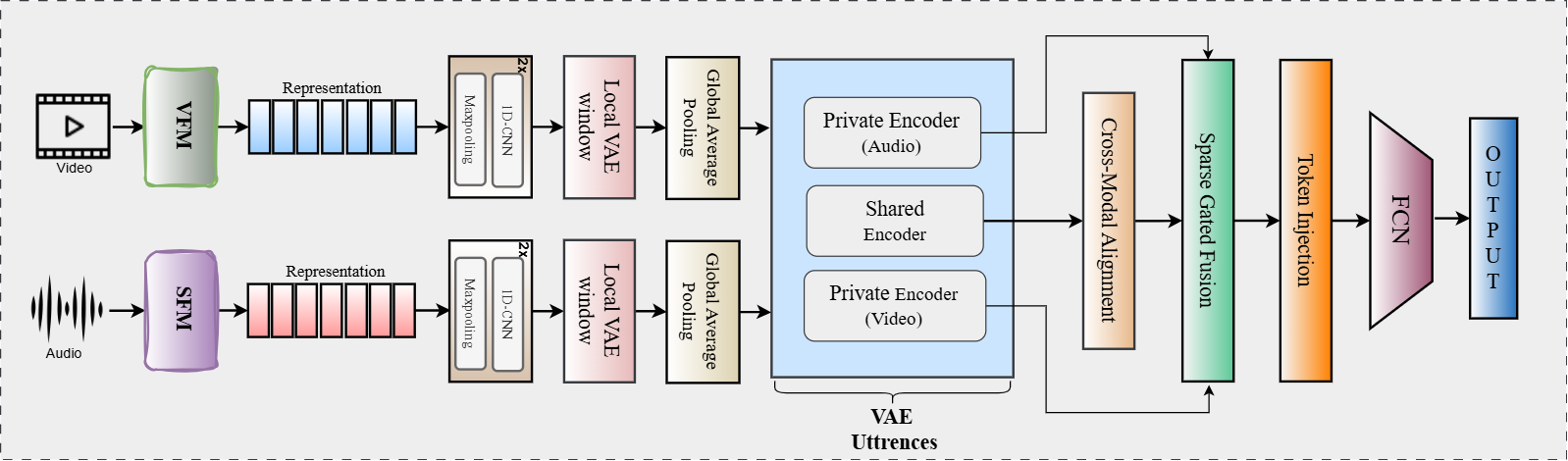}
   \caption{Proposed modeling architecture : \textbf{DIVINE}}
\label{fig:archi}
\end{figure*}

\noindent \textbf{{DIVINE:}}
We propose \textbf{DIVINE}, a novel multimodal learning framework tailored for neuro-facial disorder assessment. It is built upon a fully disentangled pipeline that incorporates \textit{hierarchical latent modeling, gated cross-modal fusion, and clinical token-aware dense reasoning over synchronized audio and video inputs}. Intuitively, DIVINE treats speech and facial motion as two noisy views of the same latent motor process. Local VAEs compress short-time articulatory dynamics, while utterance-level VAEs factor out modality-specific noise from shared neuro-motor traits. The sparse fusion gates then balance visual and acoustic evidence, and clinical tokens encourage interpretable aggregation across modalities. The overall architecture of the proposed framework is illustrated in Figure~\ref{fig:archi}. DIVINE combines local and utterance-level VAEs to achieve (i) compression of short-time patterns, and (ii) factorization of modality-specific vs modality-shared cues. A lightweight cross-modal decoder utilizes a penalty to align representations to limit leakage between subspaces, while sparse gated fusion channels evidence from the more reliable stream. Ultimately, symptom tokens provide a compact grid that invites clinically meaningful aggregation; it is necessary to avoid over-reliance on a single strong encoder to be robust when a modality is degraded or altogether missing.

\noindent We extract foundational audio and video representations from raw inputs using frozen pretrained models. Let the raw video and audio inputs be denoted as
\[
v \in \mathbb{R}^{T_v \times H \times W \times C}, 
\quad
a \in \mathbb{R}^{T_a}.
\]
We extract frozen representations using pretrained foundation models:
\begin{equation*}
\begin{aligned}
X_v &= \mathrm{VFM}(v)\in\mathbb{R}^{T_v \ast d_v},\\
X_a &= \mathrm{SFM}(a)\in\mathbb{R}^{T_a \ast d_a}.
\end{aligned}
\end{equation*}
\paragraph{Local Temporal Refinement}
We first refine the local temporal structure for each modality using CNN-based feature transformation. For each modality \(m \in \{v, a\}\), we apply a temporal refinement stage:
\[
X'_m = \mathrm{CNN}_m(X_m) \in \mathbb{R}^{T'_m \times d'_m}.
\]
where \(\mathrm{CNN}_m\) consists of a 1D Convolution, Batch Normalization, ReLU activation, and Max Pooling.

\paragraph{Local VAE (VAE\_window)}
We apply a local VAE over temporally refined segments. For each temporal index \(t = 1, \dots, T''\) and modality \(m \in \{v, a\}\), the local variational encoding and decoding steps are:
\begin{equation*}
\begin{gathered}
(\mu_w^m(t),\, \log\sigma_w^m(t)) = f^{w}_{\mathrm{enc}}\!\big(X'_m[t]\big), \\
z_{\mathrm{sig}}^{m}(t) = \mu_w^m(t)
  + \exp\!\left(\tfrac{1}{2}\log\sigma_w^m(t)\right)\odot \epsilon, \\
\epsilon \sim \mathcal{N}(0, I), \\
\hat X'_m[t] = f^{w}_{\mathrm{dec}}\!\big(z_{\mathrm{sig}}^{m}(t)\big)
\end{gathered}
\end{equation*}

\noindent The local VAE loss is defined as:
\[
\begin{aligned}
\mathcal{L}^m_{w} &=
\frac{1}{T''} \sum_{t=1}^{T''} 
\big\| X'_m[t] - \hat{X}'_m[t] \big\|^2 \\
&\quad + 
\mathrm{KL}\!\left(
\mathcal{N}\!\big(\mu^m_w(t), (\sigma^m_w(t))^2\big)
\,\|\, 
\mathcal{N}(0, I)
\right)
\end{aligned}
\]

\paragraph{Global Average Pooling}
We summarize local latent variables across time via global average pooling to obtain fixed-length utterance-level embeddings.
\[
\bar{z}^m = \frac{1}{T''} \sum_{t=1}^{T''} z^m_{\mathrm{sig}}(t) \in \mathbb{R}^{d_w}
\]
\paragraph{Utterance-Level VAE (VAE\_utterance)}

We disentangle modality-invariant (shared) and modality-specific (private) representations at the utterance level using two parallel variational autoencoders (VAEs). For each modality \( m \in \{v, a\} \), the shared encoder is weight-tied across modalities and maps the global latent representation \( \bar{z}^m \) to the parameters of a Gaussian distribution, producing a mean \( \mu^m_s \) and log-variance \( \log\sigma^m_s \). A shared latent variable is sampled using the reparameterization trick as
\[
z^m_{\mathrm{shared}} = \mu^m_s 
+ \exp\!\left(\tfrac{1}{2}\log\sigma^m_s\right)\odot \epsilon, 
\quad 
\epsilon \sim \mathcal{N}(0, I)
\]

In parallel, a modality-specific private encoder \( f^{p,m}_{\mathrm{enc}} \), which is unique to each modality, generates the private latent representation by producing \( \mu^m_p \) and \( \log\sigma^m_p \), from which the private vector is sampled as
\[
z^m_{\mathrm{priv}} = \mu^m_p 
+ \exp\!\left(\tfrac{1}{2}\log\sigma^m_p\right)\odot \epsilon
\]

To regularize shared and private encodings, we define the utterance-level VAE loss as the sum of a reconstruction term and KL divergence penalties. The total loss is represented as:
\[
\begin{aligned}
\mathcal{L}^m_{u} &= \mathcal{L}^m_{\mathrm{rec}} 
+ \beta_s\,\mathrm{KL}\!\left(
\mathcal{N}\!\big(\mu^m_s, (\sigma^m_s)^2\big)
\,\|\, 
\mathcal{N}(0, I)
\right) \\
&\quad + 
\beta_p\,\mathrm{KL}\!\left(
\mathcal{N}\!\big(\mu^m_p, (\sigma^m_p)^2\big)
\,\|\, 
\mathcal{N}(0, I)
\right)
\end{aligned}
\]

where \(\beta_s\) and \(\beta_p\) are hyperparameters controlling the relative importance of the shared and private KL divergence terms.
\begin{table*}[!hbt]
\setlength{\tabcolsep}{7pt}
\centering
\scriptsize
\begin{tabular}{l|cccc|cccc|cccc}
  \toprule
  \multicolumn{1}{l|}{\textbf{Combinations}}
    & \multicolumn{4}{c|}{\textbf{Speech Videos}}
    & \multicolumn{4}{c|}{\textbf{Testing Only Video}}
    & \multicolumn{4}{c}{\textbf{Testing Only Audio}} \\
  \cmidrule(lr){2-5} \cmidrule(lr){6-9} \cmidrule(lr){10-13}
  \multicolumn{1}{l|}{}
    & \textbf{A \(\uparrow\)} & \textbf{F1 \(\uparrow\)} & \textbf{R \(\downarrow\)} & \textbf{M \(\downarrow\)}
    & \textbf{A \(\uparrow\)} & \textbf{F1 \(\uparrow\)} & \textbf{R \(\downarrow\)} & \textbf{M \(\downarrow\)}
    & \textbf{A \(\uparrow\)} & \textbf{F1 \(\uparrow\)} & \textbf{R \(\downarrow\)} & \textbf{M \(\downarrow\)} \\
  \midrule
\multicolumn{13}{c}{\textbf{Concatenation}} \\
\midrule
\textbf{Vi + WV}   & 84.55 & 83.64 & 4.82 & 3.96
                   & 79.25 & 79.11 & 11.72 & 10.27
                   & 74.46 & 73.61 & 11.66 & 10.29 \\
\textbf{Vi + W2}   & 83.41 & 82.61 & 4.86 & 3.74
                   & 78.73 & 77.79 & 12.25 & 9.69
                   & 72.31 & 71.55 & 12.13 & 9.65 \\
\textbf{Vi + WR}   & 87.25 & 86.23 & 4.75 & 3.87
                   & 79.22 & 78.75 & 11.55 & 9.91
                   & 78.20 & 77.42 & 11.68 & 10.04 \\
\textbf{Vi + XV}   & 91.64 & 90.85 & 4.68 & 3.91
                   & 80.68 & 79.68 & 12.13 & 9.52
                   & 81.16 & 80.29 & 12.31 & 10.55 \\
\textbf{Vi + HT}   & 85.32 & 84.64 & 4.80 & 3.98
                   & 78.41 & 77.79 & 12.20 & 9.60
                   & 74.66 & 73.90 & 12.10 & 10.55 \\
\textbf{Vi + TR}   & 92.65 & 91.11 & 4.29 & 3.52
                   & 80.65 & 79.23 & 11.61 & 9.88
                   & \textbf{82.27} & \textbf{81.44} & 10.52 & 8.78 \\
\midrule
\textbf{V2 + WV}   & 86.36 & 85.29 & 4.86 & 3.49
                   & 83.08 & 82.17 & 11.96 & 10.59
                   & 72.61 & 71.81 & 11.89 & 9.71 \\
\textbf{V2 + W2}   & 85.27 & 84.56 & 4.81 & 3.45
                   & 82.18 & 81.16 & 11.72 & 9.74
                   & 73.28 & 72.27 & 11.86 & 9.62 \\
\textbf{V2 + WR}   & 87.21 & 86.21 & 4.67 & 3.38
                   & 83.52 & 82.39 & 11.70 & 10.46
                   & 74.63 & 73.80 & 11.84 & 9.39 \\
\textbf{V2 + XV}   & \underline{93.22} & \underline{92.55} & \textbf{3.72} & 2.75
                   & 83.34 & 82.51 & 11.09 & 10.03
                   & 80.04 & 79.21 & \underline{10.09} & \textbf{8.15} \\
\textbf{V2 + HT}   & 87.65 & 86.08 & 4.78 & 3.42
                   & 83.21 & 82.65 & 11.65 & \textbf{8.39}
                   & 74.95 & 74.10 & 11.82 & 9.35 \\
\textbf{V2 + TR}   & 90.99 & 89.24 & 3.76 & \underline{2.69}
                   & 83.54 & 82.08 & 11.65 & 9.87
                   & 81.88 & 81.01 & 10.31 & 9.61 \\
\midrule
\textbf{VV + WV}& 82.19 & 81.65 & 6.29 & 5.16
                   & 77.69 & 79.11 & 13.44 & 10.81
                   & 72.09 & 71.38 & 13.52 & 11.43 \\
\textbf{VV + W2}& 81.54 & 79.69 & 6.23 & 4.77
                   & 76.82 & 75.17 & 13.19 & 10.58
                   & 71.22 & 70.11 & 13.66 & 11.59 \\
\textbf{VV + WR}& 85.47 & 84.43 & 6.39 & 5.23
                   & 78.23 & 76.86 & 13.39 & 10.73
                   & 76.21 & 75.09 & 13.99 & 11.71 \\
\textbf{VV + XV}& 89.36 & 88.14 & 6.38 & 4.29
                   & 79.28 & 78.35 & 13.25 & 10.59
                   & 79.14 & 78.16 & 13.52 & 11.29 \\
\textbf{VV + HT}& 83.17 & 82.64 & 6.85 & 4.12
                   & 77.15 & 76.38 & 13.11 & 10.34
                   & 72.68 & 71.24 & 13.25 & 11.05 \\
\textbf{VV + TR}& 90.35 & 89.15 & 6.16 & 4.85
                   & 78.61 & 77.29 & 12.05 & 10.27
                   & 81.53 & 80.17 & 11.23 & 9.28 \\
\midrule
\textbf{DS + WV}   & 91.58 & 90.09 & 4.59 & 3.36
                   & 86.08 & 85.23 & \underline{11.03} & 9.15
                   & 74.28 & 73.46 & 10.93 & \underline{8.20} \\
\textbf{DS + W2}   & 89.25 & 88.34 & 4.52 & 3.23
                   & 85.56 & 84.27 & 11.49 & \underline{9.04}
                   & 72.44 & 71.66 & 11.42 & 10.23 \\
\textbf{DS + WR}   & 92.66 & 91.01 & 4.36 & 3.08
                   & \underline{86.09} & \textbf{85.37} & 11.31 & 10.70
                   & 78.34 & 77.51 & 11.40 & 10.64 \\
\textbf{DS + XV}   & 92.69 & 91.14 & 3.89 & 2.77
                   & 84.53 & 83.20 & \textbf{10.01} & 9.70
                   & 80.10 & 79.29 & \textbf{10.07} & 9.32 \\
\textbf{DS + HT}   & 90.27 & 89.64 & 4.47 & 3.15
                   & 85.88 & 85.17 & 11.27 & 9.99
                   & 74.83 & 74.03 & 11.37 & 10.11 \\
\textbf{DS + TR}   & \textbf{94.65} & \underline{93.87} & \underline{3.73} & \textbf{2.61}
                   & \textbf{86.33} & \underline{85.27} & 12.06 & 10.10
                   & 82.01 & 81.15 & 10.12 & 9.19 \\
\midrule
\textbf{KI + WV}   & 81.63 & 80.52 & 5.98 & 4.78
                   & 72.07 & 70.61 & 14.70 & 12.37
                   & 72.58 & 71.76 & 14.84 & 13.44 \\
\textbf{KI + W2}   & 79.64 & 78.11 & 5.91 & 4.70
                   & 71.96 & 70.55 & 14.35 & 12.01
                   & 74.89 & 74.09 & 14.25 & 12.92 \\
\textbf{KI + WR}   & 84.25 & 83.64 & 5.92 & 4.69
                   & 72.25 & 70.55 & 14.64 & 12.10
                   & 74.71 & 73.88 & 14.52 & 13.04 \\
\textbf{KI + XV}   & 85.66 & 84.29 & 5.24 & 4.16
                   & 74.20 & 72.92 & 12.88 & 10.14
                   & \underline{82.05} & \underline{81.22} & 13.01 & 12.09 \\
\textbf{KI + HT}   & 80.56 & 79.65 & 5.85 &
4.62 & 71.48 & 70.76 & 14.76 & 11.98 & 80.13 & 79.30 & 14.71 & 13.12 \\
\textbf{KI + TR}   & 86.19 & 85.35 & 5.17 & 4.04
                   & 85.34 & 84.28 & 12.85 & 10.14
                   & 75.39 & 74.25 & 12.94 & 11.14 \\
\midrule
\textbf{TA + WV}   & 82.26 & 81.93 & 5.66 & 4.38
                   & 74.35 & 73.62 & 14.25 & 10.55
                   & 72.36 & 71.54 & 14.44 & 13.55 \\
\textbf{TA + W2}   & 80.52 & 79.67 & 5.43 & 4.27
                   & 74.55 & 73.64 & 13.59 & 10.42
                   & 74.72 & 73.89 & 13.78 & 12.53 \\
\textbf{TA + WR}   & 83.15 & 82.65 & 5.76 & 4.51
                   & 74.56 & 73.62 & 13.88 & 11.12
                   & 74.54 & 73.73 & 14.07 & 13.08 \\
\textbf{TA + XV}   & 86.74 & 85.51 & 5.11 & 4.01
                   & 76.77 & 75.91 & 13.07 & 10.36
                   & 81.96 & 81.12 & 13.17 & 12.51 \\
\textbf{TA + HT}   & 82.12 & 81.68 & 5.49 & 4.23
                   & 75.39 & 75.25 & 13.40 & 10.23
                   & 80.07 & 79.24 & 13.56 & 12.33 \\
\textbf{TA + TR}   & 90.52 & 89.58 & 5.05 & 3.86
                   & 77.15 & 75.84 & 12.58 & 10.73
                   & 78.15 & 77.33 & 12.57 & 9.75 \\
\bottomrule
\end{tabular}
\caption{Performance on combinations of VFM and SFM on simple concatenation combinations across three settings: speech videos, video-only, and audio-only. All scores are reported in percentage (\%) and averaged over 5-fold cross-validation.}
\label{tab:2}
\end{table*}

\paragraph{Cross-Modal Alignment}
We decode the video-shared representation into the audio-shared latent space:
\[
\begin{aligned}
\hat{z}_a &= D_a\!\left(z^v_{\mathrm{shared}}\right), \\
\mathcal{L}_{\mathrm{cycle}} &= 
\left\|\hat{z}_a - z^a_{\mathrm{shared}}\right\|_2^2
\end{aligned}
\]

\paragraph{Sparse Gated Fusion}
We compute a sparse, learnable fusion of modality-specific and shared embeddings to dynamically weigh audio and video cues.
\[
\begin{aligned}
g_v &= \sigma\!\left(W_v z^v_{\mathrm{priv}} + b_v\right), \\
g_a &= \sigma\!\left(W_a z^a_{\mathrm{priv}} + b_a\right)
\end{aligned}
\]

\noindent The fused latent representation is computed as:
\[
h_{\mathrm{fused}} = 
g_v \odot z^v_{\mathrm{shared}} 
+ g_a \odot z^a_{\mathrm{shared}} 
\in \mathbb{R}^{d_s}.
\]
\[
\mathcal{L}_{\mathrm{sparse}} = 
\|g_v\|_1 + \|g_a\|_1
\]

\paragraph{Token Injection and Dense Layer}
Let \(T_1, \dots, T_K \in \mathbb{R}^{d_s}\) be learnable clinical symptom tokens. Concatenate and input to dense layer:
\[
S = [T_1, \dots, T_K, h_{\mathrm{fused}}] 
\in \mathbb{R}^{(K+1) \times d_s},
\]
\[
H_{\mathrm{out}} = \mathrm{Dense}(S), 
\quad 
H_{\mathrm{out}} \in \mathbb{R}^{(K+1) \times d_s}
\]

\noindent Add token specialization regularization term \(\mathcal{L}_{\mathrm{token}}\).
\paragraph{Output Heads}
Finally, we derive diagnosis and severity predictions from the fused representation using softmax or linear heads.
Let \(\mathbf{h} = H_{\mathrm{out}}[K+1]\) denote the fused output. The classification and severity predictions are:
\[
\begin{aligned}
\hat{y}_{\mathrm{cls}} &= 
\mathrm{softmax}\!\left(W_{\mathrm{cls}}\mathbf{h} + b_{\mathrm{cls}}\right), \\
\hat{y}_{\mathrm{sev}} &= 
\mathrm{softmax}\!\left(W_{\mathrm{sev}}\mathbf{h} + b_{\mathrm{sev}}\right)
\end{aligned}
\]

\noindent All non-frozen parameters are optimized end-to-end using the Adam optimizer with early stopping.
\paragraph{Joint Loss Function}
The function combines classification, severity, reconstruction, and regularization terms:
\begin{equation*}
\begin{gathered}
\mathcal{L}_{\mathrm{total}} =
\mathcal{L}_{\mathrm{cls}}
+ \alpha\,\mathcal{L}_{\mathrm{sev}}
+ \epsilon\Big(
\mathcal{L}_{\mathrm{cycle}}
+ \mathcal{L}_{\mathrm{sparse}} \\
+ \epsilon\,\lambda\,\mathcal{L}_{\mathrm{token}}\Big)
+ \sum_{m \in \{v, a\}}
\Big(
\mathcal{L}^m_{w} + \mathcal{L}^m_{u}
\Big)
\end{gathered}
\end{equation*}

\begin{table*}[!h]
\setlength{\tabcolsep}{7pt}
\centering
\scriptsize
\begin{tabular}{l|cccc|cccc|cccc}
  \toprule
  \multicolumn{1}{l|}{\textbf{Combinations}}
    & \multicolumn{4}{c|}{\textbf{Speech Videos}}
    & \multicolumn{4}{c|}{\textbf{Testing Only Video}}
    & \multicolumn{4}{c}{\textbf{Testing Only Audio}} \\
  \cmidrule(lr){2-5} \cmidrule(lr){6-9} \cmidrule(lr){10-13}
  \multicolumn{1}{l|}{}
    & \textbf{A \(\uparrow\)} & \textbf{F1 \(\uparrow\)} & \textbf{R \(\downarrow\)} & \textbf{M \(\downarrow\)}
    & \textbf{A \(\uparrow\)} & \textbf{F1 \(\uparrow\)} & \textbf{R \(\downarrow\)} & \textbf{M \(\downarrow\)}
    & \textbf{A \(\uparrow\)} & \textbf{F1 \(\uparrow\)} & \textbf{R \(\downarrow\)} & \textbf{M \(\downarrow\)} \\
  \midrule
  \multicolumn{13}{c}{\textbf{\texttt{DIVINE}}} \\
  \midrule
  \textbf{Vi + WV}  & 86.99 & 86.11 & 2.60 & 2.10 & 81.69 & 82.26 & 6.32 & 5.31 & 76.54 & 75.43 & 6.13 & 5.09 \\
  \textbf{Vi + W2}  & 85.23 & 84.39 & 2.89 & 1.92 & 81.20 & 80.75 & 6.64 & 4.92 & 74.79 & 73.90 & 6.08 & 4.73 \\
  \textbf{Vi + WR}  & 89.45 & 88.55 & 2.84 & 1.95 & 81.75 & 81.42 & 6.25 & 5.19 & 80.55 & 79.54 & 6.06 & 5.02 \\
  \textbf{Vi + XV}  & 93.06 & 92.17 & 2.47 & 2.11 & 83.77 & 82.35 & 6.08 & 5.20 & 83.67 & 82.65 & 5.90 & 4.89 \\
  \textbf{Vi + HT}  & 87.25 & 86.42 & 2.81 & 2.33 & 80.98 & 80.55 & 6.37 & 5.23 & 76.63 & 75.76 & 6.04 & 5.13 \\
  \textbf{Vi + TR}  & 94.51 & 93.63 & 2.41 & 1.78 & 83.38 & 81.82 & 5.69 & 4.26 & 83.61 & 82.06 & 5.51 & 4.43 \\
  \midrule
  \textbf{V2 + WV} & 88.04 & 87.23 & 2.88 & 1.76 & 85.89 & 84.87 & 6.47 & 4.67 & 74.75 & 73.97 & 6.31 & 4.47 \\
  \textbf{V2 + W2} & 88.54 & 84.68 & 2.59 & 2.01 & 84.95 & 83.99 & 6.27 & 4.53 & 75.30 & 74.36 & 6.22 & 4.21 \\
  \textbf{V2 + WR} & 89.48 & 88.59 & 2.35 & 1.75 & 86.51 & 85.64 & 6.19 & 4.54 & 76.98 & 75.91 & 5.86 & 4.31 \\
 \textbf{V2 + XV} & \cellcolor{green!25}\textbf{96.41} & \cellcolor{green!25}\textbf{95.68} & \cellcolor{yellow!25}\textbf{2.16} & \cellcolor{yellow!25}\textbf{1.51} & 86.59 & 85.48 & \cellcolor{blue!25}\textbf{4.95} & 3.71 & \cellcolor{yellow!25}\textbf{83.71} & \cellcolor{yellow!25}\textbf{82.27} & \cellcolor{blue!25}\textbf{4.68} & \cellcolor{yellow!25}\textbf{3.45} \\

  \textbf{V2 + HT} & 89.85 & 88.97 & 2.58 & 2.04 & 86.26 & 85.91 & 6.36 & 4.51 & 77.07 & 76.22 & 6.06 & 4.29 \\
  \textbf{V2 + TR} & 95.16 & 94.68 & \cellcolor{green!25}\textbf{2.08} & \cellcolor{green!25}\textbf{1.39} & 86.93 & 84.83 & \cellcolor{yellow!25}\textbf{5.06} & \cellcolor{green!25}\textbf{3.50} & \cellcolor{green!25}\textbf{84.03} & \cellcolor{green!25}\textbf{83.08} & \cellcolor{yellow!25}\textbf{4.84} & \cellcolor{green!25}\textbf{3.41} \\
  \midrule
  \textbf{VV + WV} & 85.48 & 84.60 & 2.71 & 2.21 & 79.94 & 80.61 & 8.43 & 6.90 & 74.22 & 73.46 & 8.11 & 6.60 \\
  \textbf{VV + W2} & 83.72 & 82.81 & 3.09 & 2.03 & 79.62 & 78.57 & 8.27 & 6.32 & 73.21 & 72.08 & 7.85 & 6.14 \\
  \textbf{VV + WR} & 87.89 & 87.04 & 3.03 & 2.06 & 79.62 & 78.88 & 8.46 & 6.95 & 78.29 & 77.14 & 8.30 & 6.58 \\
  \textbf{VV + XV} & 91.35 & 90.32 & 2.64 & 2.24 & 81.33 & 79.88 & 8.33 & 5.64 & 81.44 & 80.32 & 8.03 & 5.38 \\
  \textbf{VV + HT} & 86.11 & 85.27 & 2.99 & 2.46 & 78.73 & 78.13 & 8.97 & 5.53 & 74.66 & 73.06 & 8.61 & 5.25 \\
  \textbf{VV + TR} & 93.47 & 92.49 & 2.50 & 1.86 & 81.75 & 79.99 & 8.02 & 6.53 & 82.24 & 81.31 & 7.90 & 6.23 \\
  \midrule
  \textbf{DS + WV} & 93.83 & 92.91 & 2.47 & 1.97 & 88.64 & 87.49 & 6.14 & 4.37 & 76.45 & 75.47 & 5.75 & 4.22 \\
  \textbf{DS + W2} & 91.11 & 90.28 & 2.52 & 1.84 & 88.55 & 87.45 & 6.05 & 4.21 & 74.72 & 73.78 & 5.70 & 4.05 \\
  \textbf{DS + WR} & 95.48 & 94.55 & 2.49 & 1.74 & \cellcolor{green!25}\textbf{89.01} & \cellcolor{green!25}\textbf{88.66} & 5.74 & 4.02 & 80.71 & 79.78 & 5.57 & 3.89 \\
  \textbf{DS + XV} & \cellcolor{yellow!25}\textbf{95.56} & \cellcolor{yellow!25}\textbf{94.63} & 2.26 & 1.61 & \cellcolor{yellow!25}\textbf{88.85} & \cellcolor{yellow!25}\textbf{88.02} & 5.25 & \cellcolor{yellow!25}\textbf{3.61} & 82.24 & 81.29 & 5.06 & 3.49 \\
  \textbf{DS + HT} & 92.99 & 92.11 & 2.55 & 1.72 & 87.89 & 86.53 & 5.88 & 4.49 & 76.97 & 76.13 & 5.81 & 4.32 \\
  \textbf{DS + TR} & \cellcolor{blue!25}\textbf{98.26} & \cellcolor{blue!25}\textbf{97.51} & \cellcolor{blue!25}\textbf{1.93} & \cellcolor{blue!25}\textbf{1.12} & \cellcolor{blue!25}\textbf{89.27} & \cellcolor{blue!25}\textbf{88.23} & \cellcolor{green!25}\textbf{5.02} & \cellcolor{blue!25}\textbf{3.44} & \cellcolor{blue!25}\textbf{84.34} & \cellcolor{blue!25}\textbf{83.20} & \cellcolor{green!25}\textbf{4.80} & \cellcolor{blue!25}\textbf{3.31} \\
  \midrule
  \textbf{KI + WV} & 83.26 & 82.41 & 3.43 & 2.55 & 75.11 & 73.26 & 8.02 & 6.39 & 74.77 & 73.91 & 7.62 & 6.08 \\
  \textbf{KI + W2} & 81.22 & 80.45 & 3.53 & 2.39 & 74.94 & 72.97 & 7.94 & 6.25 & 77.04 & 76.21 & 7.62 & 5.90 \\
  \textbf{KI + WR} & 86.07 & 85.23 & 3.37 & 2.57 & 75.26 & 73.29 & 7.80 & 6.24 & 76.86 & 75.90 & 7.55 & 5.88 \\
  \textbf{KI + XV} & 87.19 & 86.28 & 2.79 & 2.31 & 76.99 & 75.85 & 6.30 & 5.42 & 82.23 & 81.36 & 6.71 & 5.37 \\
  \textbf{KI + HT} & 82.14 & 81.37 & 3.44 & 2.54 & 73.94 & 73.16 & 7.77 & 6.23 & 82.26 & 81.33 & 7.54 & 5.97 \\
  \textbf{KI + TR} & 88.39 & 87.63 & 2.94 & 2.38 & 88.25 & 87.30 & 6.82 & 5.40 & 77.95 & 76.56 & 6.47 & 5.21 \\
  \midrule
  \textbf{TA + WV} & 84.97 & 84.13 & 3.35 & 2.54 & 76.80 & 76.44 & 7.40 & 5.85 & 74.57 & 73.63 & 7.36 & 5.58 \\
  \textbf{TA + W2} & 82.88 & 82.07 & 2.93 & 2.17 & 76.96 & 76.42 & 7.07 & 5.61 & 76.94 & 75.98 & 6.94 & 5.47 \\
  \textbf{TA + WR} & 85.99 & 85.15 & 3.43 & 2.56 & 77.01 & 76.39 & 7.76 & 5.92 & 76.77 & 75.85 & 7.38 & 5.65 \\
  \textbf{TA + XV} & 88.63 & 87.78 & 2.57 & 2.10 & 79.24 & 78.75 & 6.72 & 5.29 & 82.10 & 81.22 & 6.50 & 5.14 \\
  \textbf{TA + HT} & 84.35 & 83.48 & 2.78 & 2.16 & 77.91 & 78.08 & 7.41 & 5.63 & 81.19 & 80.30 & 7.13 & 5.48 \\
  \textbf{TA + TR} & 93.20 & 92.32 & 2.75 & 2.24 & 80.06 & 78.71 & 6.75 & 5.19 & 80.43 & 79.48 & 6.52 & 4.84 \\
\bottomrule
\end{tabular}
\caption{Performance on combinations of VFM and SFM on \textbf{\texttt{DIVINE}} framework across three settings: speech videos, video-only, and audio-only. All values are reported in percentage (\%) and averaged over 5-fold cross-validation.}
\label{tab:3}
\end{table*}
\section{Experiments}
\label{Exp-result}
\noindent \textbf{Benchmark Dataset:}  We conduct our experiments on the Toronto NeuroFace (TNF) dataset~\cite{bandini2020new}, which contains synchronized audio and video recordings from cognitively intact adults across three clinical groups: ALS, stroke, and healthy controls. We use subject-wise 5-fold cross-validation: folds are defined over participants, and all recordings from a participant appear in only one fold split (train/val/test) per run, preventing speaker leakage across splits. Detailed information on the dataset, task design, and annotation procedures is provided in Appendix~\ref{corpus},\ref{corpuspreprocesss}. \newline
\noindent \textbf{Training Details:} We use softmax activation in the output layers for both classification and severity prediction heads to produce probability distributions. All models are trained using the Adam optimizer with a learning rate of $10^{-3}$, a batch size 32, and categorical cross-entropy loss. Training is performed for 50 epochs with early stopping and dropout regularization to mitigate overfitting. For all \textbf{DIVINE} experiments, we fix the hyperparameters: $\alpha = 2$, $\epsilon = 0.1$, and $\lambda = 0.4$, selected based on preliminary validation performance. These values are kept consistent across all fusion and ablation experiments. \newline
\noindent \textbf{Experimental Results:}
\noindent Table~\ref{tab-1} shows how each Video Foundation Model (VFM) and Speech Foundation Model (SFM) performs on the TNF speech–video samples, using both FCN and CNN backbones. Among the VFMs, DeepSeek-VL2 (DS) leads with a CNN accuracy of 88.94\% and F1 of 86.57\%, and achieves the lowest regression errors (MAE = 8.58, RMSE = 6.81) as well as the lowest multitask errors (MAE = 7.58, RMSE = 9.38). VideoMAE V2 follows closely (85.69 \% accuracy, 83.58 \% F1; MAE = 8.73, RMSE = 6.84). Handcrafted kinematic and temporal features lag behind (76–79 \% accuracy with CNN), highlighting the value of pretrained vision encoders. In the audio domain, TRILLsson (TR) is top: it records 90.51 \% accuracy and 88.69 \% F1 in the multitask setting, with MAE = 6.12 and RMSE = 7.01. Wav2Vec 2.0 and Whisper also perform well (e.g. Wav2Vec 2.0 reaches 89.27 \% accuracy, 87.64 \% F1), while WavLM and X-vector show weaker regression consistency. Overall, CNN backbones outperform FCNs, confirming their strength at capturing local temporal patterns.

\noindent Next, we fuse VFMs and SFMs using a simple embedding concatenation (Table~\ref{tab:2}). Here, DS + TR achieves 94.65 \% accuracy and 93.87 \% F1 on full speech–video inputs, while still holding 86.33 \% accuracy when only video is available and 82.01 \% when only audio is available. VideoMAE V2 + X-vector also performs strongly (93.22 \% accuracy, 92.55 \% F1). These results show that even a straightforward fusion of embeddings leverages complementary modality information and degrades gracefully when one modality is unavailable.
\noindent Finally, Table~\ref{tab:3} reports our \texttt{DIVINE} disentangled fusion. The best pair, DS + TR, reaches \textbf{98.26\,\%} accuracy and \textbf{97.51\,\%} F1 when both audio and video embeddings are provided. When evaluated with only video embeddings, DS + TR still scores \textbf{89.27\,\%} accuracy (F1 = 88.23), and when evaluated with only audio embeddings, it achieves \textbf{84.34\,\%} accuracy (F1 = 83.20). Other strong pairs include VideoMAE V2 + X-vector (96.41\,\% accuracy, 95.68\,\% F1) and ViViT + TR (over 90\,\% accuracy). All results are averaged over five random seeds and five-fold cross-validation. We report mean ± std and use paired Wilcoxon tests versus the strongest baseline (p < 0.05). Results remain stable across seeds, confirming reliability. \newline
To assess DIVINE’s ability to handle purely visual input, we test on non‐speech videos \textit{(Detailed results for these experiments are presented in (Appendix~\ref{NSVS}, Tables~\ref{tab:4}–\ref{tab:6}}). Table~\ref{tab:4}, DS individually achieves 89.26 \% accuracy and 88.29 \% F1 (MAE = 6.02, RMSE = 8.06). When we simply concatenate VFM and SFM embeddings (Table~\ref{tab:5}), DS + X-vector still reaches 87.24 \% accuracy and 86.01 \% F1, showing that pre-computed audio features can aid video-only inference. With our \textbf{DIVINE} framework fusion (Table~\ref{tab:6}), DS + TR climbs to 92.58 \% accuracy and 91.63 \% F1 (MAE = 3.84, RMSE = 5.55), confirming that the model maintains strong performance using only visual information. Refer to (Appendix~\ref{NSVS}) for more detail. Additionally, we also present confusion matrices of key configurations in Figure~\ref{fig:cm_8} (Appendix~\ref{CM}).

\subsection{Ablation Study}
To assess the contribution of key components in the proposed framework, we conduct a detailed ablation study along three axes: 

\subsubsection{Role of Modalities}
While unimodal performance was previously discussed in Section~\ref{Exp-result}. We revisit these results here to isolate the individual contribution of each modality. We retain the full model but remove the audio or video input at inference time.

\begin{table}[!h]
\centering
\scriptsize
\setlength{\tabcolsep}{8pt}
\renewcommand{\arraystretch}{1.2}
\begin{tabular}{l|c|c|c|c}
\toprule
\textbf{Setting} & \textbf{A} $\uparrow$ & \textbf{F1} $\uparrow$ & \textbf{M} $\downarrow$ & \textbf{R} $\downarrow$ \\
\midrule
\textbf{DIVINE} (Audio + Video) & \textbf{98.26} & \textbf{97.51} & \textbf{1.12} & \textbf{1.93} \\
Audio only             & 89.27          & 88.23          & 5.02          & 3.44 \\
Video only             & 84.34          & 83.20          & 4.80          & 3.31 \\
\bottomrule
\end{tabular}
\caption{Performance representing the role of modality.}
\label{tab:ablation1}
\end{table}
\subsubsection{Role of Regularization}
We compare the three regularization components in DIVINE: Cycle-consistency (CC) loss, sparse gating (SG), and token reconstruction (TR) loss. Each component is removed independently to evaluate its influence on performance.

\begin{table}[!h]
\centering
\scriptsize
\setlength{\tabcolsep}{8pt}
\renewcommand{\arraystretch}{1.2}
\begin{tabular}{l|c|c|c|c}
\toprule
\textbf{Configuration} & \textbf{A} $\uparrow$ & \textbf{F1} $\uparrow$ & \textbf{M} $\downarrow$ & \textbf{R} $\downarrow$ \\
\midrule
\textbf{DIVINE}             & \textbf{98.26} & \textbf{97.51} & \textbf{1.12} & \textbf{1.93} \\
w/o Cycle Consistency     & 96.14          & 94.95          & 1.68          & 2.37 \\
w/o Sparse Gating         & 95.83          & 94.21          & 1.84          & 2.65 \\
w/o Token Reconstruction  & 95.62          & 93.89          & 1.90          & 2.71 \\
\bottomrule
\end{tabular}
\caption{Performance representing the role of regularization.}
\label{tab:ablation2}
\end{table}
\vspace{-0.2cm}
\subsubsection{Role of Latent Space Disentanglement}
DIVINE is built on disentangled representations using separate modality-invariant and modality-specific latent spaces. We compare this design against simpler variants: \textbf{Flat Fusion} and \textbf{Single-Level Latent}.

\begin{table}[!h]
\centering
\scriptsize
\setlength{\tabcolsep}{8pt}
\renewcommand{\arraystretch}{1.2}
\begin{tabular}{l|c|c|c|c}
\toprule
\textbf{Architecture Variant} & \textbf{A} $\uparrow$ & \textbf{F1} $\uparrow$ & \textbf{M} $\downarrow$ & \textbf{R} $\downarrow$ \\
\midrule
\textbf{DIVIN}E (2-Level VAE)      & \textbf{98.26} & \textbf{97.51} & \textbf{1.12} & \textbf{1.93} \\
Flat Fusion (No Bottleneck) & 93.87       & 92.10          & 2.11          & 2.88 \\
Single-Level Latent Fusion  & 95.22       & 93.80          & 1.85          & 2.62 \\
\bottomrule
\end{tabular}
\caption{Performance representing the role of subspace disentanglement.}
\label{tab:ablation3}
\end{table}

\section{Conclusion}
In conclusion, we introduced \textbf{DIVINE}, a disentangled multimodal framework for joint classification and severity estimation of neuro-facial disorders. The approach is built upon hierarchical latent modeling, sparse gated fusion, and learnable symptom tokens, enabling effective disentanglement and integration of clinical cues from orofacial video and speech modalities. We conduct extensive experiments on the Toronto NeuroFace dataset across speech and non-speech tasks, unimodal and multimodal conditions, and scenarios with missing modalities. Performance demonstrates that our framework consistently outperforms individual audio/video models and baseline fusion techniques. Notably, the concatenation of DeepSeek-VL2 and TRILLsson through \textbf{DIVINE} achieves SOTA performance.
\subsection*{Limitations and Future Work}
While our extensive in‐domain evaluation on TNF demonstrates DIVINE’s strong performance, full cross‐dataset validation is contingent on access to suitable external corpora. 
In the camera‐ready version, we plan—subject to data availability—to evaluate our audio and video encoders separately on external unimodal benchmarks, since no suitable corpus provides both synchronized audio–video recordings. 
\subsection*{Ethical Statement}
This study uses non-public clinical data accessed with appropriate institutional approvals and participant consent. All recordings were anonymized to ensure privacy. The proposed framework is intended for research purposes and is not clinically validated for diagnostic use. 
\subsection*{Societal Impact}
DIVINE is intended as a decision-support tool for research and clinical screening workflows, not as a stand-alone diagnostic system. Predictions may be biased by recording conditions, demographic factors, or distribution shifts across sites, and erroneous outputs could lead to inappropriate clinical actions if used without safeguards. Any deployment would therefore require clinician oversight, clear communication of uncertainty, and human-in-the-loop verification. We also caution against misuse for automated triage or eligibility decisions without transparent auditing and continuous monitoring.

\bibliography{custom}
\appendix

\section{Appendix}
\label{sec:appendix}
\noindent In the Appendix, we provide: \newline
\noindent Section \ref{PTMS}: Detailed Information of Pre-trained Models. \newline
\noindent Section \ref{corpus}: Benchmark Dataset. \newline
\noindent Section \ref{corpuspreprocesss}: Data Preprocessing. \newline
\noindent Section \ref{hyperparameter}: Hyperparameters and System Configurations. \newline
\noindent Section \ref{NSVS}: Result on Non-Speech Video Samples. \newline
\noindent Section \ref{VA}: Visualization Analysis. \newline

\subsection{Detailed Information of Pre-trained Models}
\label{PTMS}
In this section, we detail the pretrained encoders used in our study. We employ pretrained speech models covering self-supervised, supervised, and multilingual training paradigms. All models are used in a frozen setting to extract utterance-level acoustic representations. \newline

\noindent \textbf{Speech Foundation Models} \newline

\noindent \textbf{WavLM}~\cite{chen2022wavlm}\footnote{\url{https://huggingface.co/microsoft/wavlm-base}}: is a self-supervised speech representation model designed to support full-stack speech processing. It is pretrained using a masked prediction and denoising objective over a diverse 94k-hour dataset composed of public English corpora. \newline
\noindent \textbf{Wav2Vec2.0}~\cite{baevski2020wav2vec}\footnote{\url{https://huggingface.co/facebook/wav2vec2-base}}: learns contextualized speech representations via contrastive prediction in the latent space. It combines a convolutional encoder with a Transformer network, masking parts of the input and optimizing discrimination against negative samples. \newline
\noindent \textbf{Whisper}~\cite{radford2023robust}\footnote{\url{https://huggingface.co/openai/whisper-base}}: is a multilingual encoder-decoder model pretrained on 680k hours of weakly labeled internet audio for transcription, translation, and speech activity detection. We use the encoder features from the base model. \newline
\noindent \textbf{x-vector} \cite{8461375}\footnote{\url{https://huggingface.co/speechbrain/spkrec-xvect-voxceleb}}: is a time-delay neural network (TDNN) trained for speaker classification using the VoxCeleb dataset. The extracted vectors are speaker-discriminative and widely adopted in speaker verification and spoof detection tasks. \newline
\noindent \textbf{HuBERT} \cite{hsu2021hubert}\footnote{\url{https://huggingface.co/facebook/hubert-base-ls960}}: is a self-supervised speech representation model that combines masked prediction with offline k-means clustering. Pretrained on large-scale datasets (e.g., LibriSpeech 960h, Libri-Light 60k), it performs state-of-the-art speech recognition and paralinguistic tasks. It is available in multiple configurations (BASE, LARGE, X-LARGE), and we use the BASE variant in frozen mode for extracting utterance-level embeddings. \newline
\noindent \textbf{TRILLsson} \cite{shor2022trillsson}\footnote{\url{https://www.kaggle.com/models/google/trillsson}}: is a lightweight self-supervised speech model designed specifically for paralinguistic speech tasks, such as emotion recognition, speaker identification, and synthetic speech detection. It is created using knowledge distillation from the CAP12 Conformer model, which was trained on 900K hours of YouTube speech data. It was trained on 58K hours of public speech data from Libri-Light and AudioSet. \newline

\noindent \textbf{Vision Foundation Models} \newline

\noindent \textbf{Video-MAE}~\cite{tong2022videomae}\footnote{\url{https://huggingface.co/docs/transformers/en/model_doc/videomae}}: follows a masked autoencoding strategy with high masking ratios (90–95\%) applied to spatiotemporal cubes. A vanilla ViT backbone is used as the encoder, and the model is trained using reconstruction as a self-supervised pretext task. \newline
\noindent \textbf{VideoMAE V2}~\cite{wang2023videomae}\footnote{\url{https://huggingface.co/OpenGVLab/VideoMAEv2-Base}}: is a scalable self-supervised video pretraining framework that extends VideoMAE with a dual masking strategy, masking both encoder and decoder tokens to reduce memory and computational load. It adopts progressive training, starting with unsupervised learning on a million-level unlabeled video corpus, followed by post-training on a labeled hybrid dataset. We employ the ViT-B variant in a frozen setting to extract clip-level facial features. \newline
\noindent \textbf{ViViT}~\cite{arnab2021vivit}\footnote{\url{https://huggingface.co/docs/transformers/en/model_doc/vivit}}: is a pure-transformer architecture that performs factorized self-attention over space and time using tubelet embeddings. We employ the ViViT-B/16×2 variant initialized from ViT image weights. \newline 
\noindent \textbf{Deepseek-VL2}~\cite{wu2024deepseek}\footnote{\url{https://github.com/deepseek-ai/DeepSeek-VL2}}: is a Mixture-of-Experts (MoE) vision-language model designed for advanced multimodal understanding. The model is trained across vision-language alignment, multimodal pretraining, and supervised fine-tuning stages on diverse tasks including visual grounding, OCR, and document understanding. Our study uses its vision encoder in a frozen mode to extract temporally aligned visual embeddings from facial video clips. \newline
\noindent \textbf{Kinematic}\footnote{\url{https://github.com/TadasBaltrusaitis/OpenFace}}: are extracted using the OpenFace 2.0 toolkit~\cite{8373812}, which provides 3D landmark positions, head pose (yaw, pitch, roll), gaze direction, and facial Action Units (AUs). \newline
\noindent \textbf{Temporal}: use a ResNet18 model pretrained on ImageNet to extract frame-level appearance embeddings. A Temporal Attention Network (TANN) is employed on top of these features to model inter-frame dependencies. \par

\subsection{Benchmark Dataset}
\label{corpus}
In this study, we used data from the Toronto NeuroFace (TNF) dataset collected by \cite{bandini2020new}, which brings together meticulously collected, high‑quality video recordings of oro‑facial gestures in healthy adults and individuals living with neurological impairment. Thirty‑six cognitively intact volunteers (11 with ALS, 14 post‑stroke, 11 age‑matched controls) each performed a battery of nine speech and non‑speech tasks---ranging from rapid syllable repetitions (``/pa/,'' ``/pa‑ta‑ka/'') and the tongue‑twister ``Buy Bobby a Puppy,'' to maximum jaw openings, lip puckers, and expressive smiles---under standardized lighting and camera distance (30--60cm, 640~$\times$~480px, $\sim$50fps). Two expert speech‑language pathologists rated every trial on symmetry, range of motion, speed, variability, and fatigue using a 5‑point scale, yielding a robust set of clinical scores (total range 5--25; inter‑rater $\kappa=0.33$--$0.61$). For over 3300 carefully chosen frames, 68 facial landmarks were hand‑annotated (inter‑rater nRMSE~=~1.36~$\pm$~0.46\%), and precise face‑bounding boxes were derived. Rich metadata---including subject demographics, task labels, video timing, and clinician ratings---is provided alongside the recordings. By combining controlled acquisition protocols with thorough ground‑truth annotations and clinical assessments. Although the dataset is not publicly available, we were granted access. To our knowledge, it is the only known resource containing synchronized, high-quality facial video and audio recordings with expert clinical annotations specific to neuro-facial disorders.

\subsection{Data Preprocessing}
\label{corpuspreprocesss}
We perform preprocessing steps to ensure data quality, consistency, and alignment across audio and video streams. For facial videos, we use the 2D Face Alignment Network (2D-FAN)~\footnote{\url{https://github.com/1adrianb/face-alignment}} to detect 68 landmarks on each frame. This helps identify the face clearly, which is visible and centrally positioned. For audio, we apply amplitude normalization and forced alignment at the utterance level using segment-level timestamps, implemented via \texttt{librosa}\cite{mcfee2025librosa} for preprocessing.

\subsection{Hyperparameters and System Configurations}
\label{hyperparameter}
The CNN architecture used for unimodal modeling begins with two 1D convolutional blocks. The first convolutional block uses 256 filters with a kernel size of 3, followed by batch normalization and max pooling (pool size = 2). The second block applies 128 filters, again with a kernel size of 3, followed by batch normalization and max pooling (pool size = 2). The flattened outputs are passed to an FCN comprising three dense layers with 256, 128, and 64 neurons, respectively, and a final task-specific output layer (either softmax or regression head). The trainable parameters for CNN models using individual pretrained representations range from 0.8 to 1.2 million, depending on the dimensionality of the extracted embeddings. This increases to 3.5–6.5 million parameters for fusion experiments due to additional transformers and fusion layers. We implement all models using the TensorFlow framework and conduct training and evaluation on an NVIDIA RTX 4050 GPU.

\subsection{Result on Non-Speech Video Samples}
\label{NSVS}
We present the complete evaluation of all VFM+SFM combinations on non‐speech video samples from the TNF dataset. Table~\ref{tab:4} reports the classification and regression metrics for each Video Foundation Model using both FCN and CNN backbones, where DeepSeek‐VL2 achieves the highest accuracy (89.26 \%) and F1 (88.29 \%). Table~\ref{tab:5} shows the results of simple embedding concatenation between VFMs and pre‐computed SFMs on video‐only inputs, demonstrating that DS + X-vector attains 87.24 \% accuracy and 86.01 \% F1 even without an audio track. Finally, Table~\ref{tab:6} provides the full results of our \texttt{DIVINE} fusion framework across all model pairings, with DS + TR leading at 92.58 \% accuracy and 91.63 \% F1 (MAE = 3.84, RMSE = 5.55). These tables offer a detailed view of model performance under purely visual conditions, complementing the concise summary in the main text.  \par
\begin{table*}[!hbt]
\setlength{\tabcolsep}{9pt}
\scriptsize
\centering
\begin{tabular}{l|cc|cc|cc|cc|cc|cc}
\toprule
\multicolumn{1}{l}{} 
  & \multicolumn{2}{c}{\textbf{FCN}} 
  & \multicolumn{2}{c}{\textbf{CNN}} 
  & \multicolumn{2}{c}{\textbf{FCN}} 
  & \multicolumn{2}{c}{\textbf{CNN}} 
  & \multicolumn{4}{c}{\textbf{CNN}} \\
\cmidrule(lr){2-3} \cmidrule(lr){4-5} \cmidrule(lr){6-7} \cmidrule(lr){8-9} \cmidrule(lr){10-13}
\multicolumn{1}{l}{} 
  & \textbf{A \(\uparrow\)} & \textbf{F1 \(\uparrow\)} 
  & \textbf{A \(\uparrow\)} & \textbf{F1 \(\uparrow\)} 
  & \textbf{M \(\downarrow\)} & \textbf{R \(\downarrow\)} 
  & \textbf{M \(\downarrow\)} & \textbf{R \(\downarrow\)} 
  & \textbf{A \(\uparrow\)} & \textbf{F1 \(\uparrow\)} & \textbf{M \(\downarrow\)} & \textbf{R \(\downarrow\)} \\
\midrule
\multicolumn{1}{l}{} 
  & \multicolumn{4}{c}{\textbf{C}} 
  & \multicolumn{4}{c}{\textbf{R}} 
  & \multicolumn{4}{c}{\textbf{M}} \\
\cmidrule(lr){2-5} \cmidrule(lr){6-9} \cmidrule(lr){10-13}
\multicolumn{13}{c}{\textbf{VFM}} \\
\midrule
\textbf{Vi}  & 81.69 & 80.25 & 83.71 & 82.05 & 7.86 & 9.84  & 7.05 & 9.12  & 84.26 & 83.31 & 8.10 & 9.38  \\
\textbf{V2}  & 82.08 & 81.67 & 86.04 & 85.22 & 6.98 & 8.82  & 6.26 & 7.59  & 87.47 & 86.09 & 7.13 & 8.62  \\
\textbf{VV}  & 79.65 & 78.16 & 81.54 & 80.46 & 9.43 & 11.29 & 8.29 & 10.65 & 83.37 & 82.09 & 9.17 & 10.99 \\
\textbf{DS}  & 86.85 & 85.98 & 89.26 & 88.29 & 6.86 & 8.35  & 6.02 & 8.06  & 90.05 & 89.84 & 6.97 & 8.76  \\
\textbf{KI}  & 73.84 & 72.26 & 75.65 & 74.57 & 8.81 & 10.57 & 7.99 & 9.73  & 76.95 & 75.21 & 8.39 & 10.51 \\
\textbf{TA}  & 78.26 & 77.23 & 80.69 & 78.08 & 7.24 & 9.83  & 7.11 & 9.29  & 80.33 & 79.62 & 7.69 & 10.03 \\
\bottomrule
\end{tabular}
\caption{Performance on video‐foundation models (VFMs) on non-speech video samples. All values are reported in percentage (\%) and averaged over 5-fold cross-validation.}
\label{tab:4}
\end{table*}

\begin{table}[!h]
\setlength{\tabcolsep}{8pt}
\centering
\scriptsize
\begin{tabular}{l|cc|cc}
  \toprule
  \textbf{Combinations} & \multicolumn{4}{c}{\textbf{Non‐speech Testing Videos}} \\
  \cmidrule(lr){2-5}
  & \textbf{A \(\uparrow\)} & \textbf{F1 \(\uparrow\)} & \textbf{R \(\downarrow\)} & \textbf{M \(\downarrow\)} \\
  \midrule
  \multicolumn{5}{c}{\textbf{VFM + SFM}} \\
  \midrule
  \textbf{Vi + WV}    & 81.49 & 80.08 & 13.23 & 11.56 \\
  \textbf{Vi + W2}    & 80.84 & 79.31 & 14.02 & 11.77 \\
  \textbf{Vi + WR}    & 80.65 & 79.44 & 13.63 & 10.89 \\
  \textbf{Vi + XV}    & 81.22 & 79.97 & 14.51 & 11.58 \\
  \textbf{Vi + HT}    & 82.39 & 81.14 & 14.77 & 11.98 \\
  \textbf{Vi + TR}    & 81.87 & 80.48 & 13.94 & 10.89 \\
  \midrule
  \textbf{V2 + WV}    & 84.52 & 83.17 & 13.55 & 11.73 \\
  \textbf{V2 + W2}    & 82.65 & 81.24 & 13.01 & 10.97 \\
  \textbf{V2 + WR}    & 83.12 & 81.96 & 13.34 & 11.67 \\
  \textbf{V2 + XV}    & 84.88 & 83.64 & 12.67 & 11.31 \\
  \textbf{V2 + HT}    & 83.99 & 82.71 & 12.58 &  9.56 \\
  \textbf{V2 + TR}    & 84.16 & 82.99 & 13.35 & 10.96 \\
  \midrule
  \textbf{VV + WV} & 80.13 & 78.58 & 15.32 & 12.36 \\
  \textbf{VV + W2} & 78.24 & 77.05 & 15.15 & 12.27 \\
  \textbf{VV + WR} & 79.05 & 77.29 & 16.34 & 12.21 \\
  \textbf{VV + XV} & 80.26 & 79.35 & 15.08 & 12.60 \\
  \textbf{VV + HT} & 81.18 & 79.31 & 16.86 & 12.85 \\
  \textbf{VV + TR} & 79.36 & 78.62 & 13.89 & 11.44 \\
  \midrule
  \textbf{DS + WV}    & 85.84 & 84.71 & 12.27 & 10.16 \\
  \textbf{DS + W2}    & 87.89 & 86.53 & 13.25 & 11.13 \\
  \textbf{DS + WR}    & 86.99 & 85.62 & 13.65 & 11.91 \\
  \textbf{DS + XV}    & 87.24 & 86.01 & 11.09 & 10.71 \\
  \textbf{DS + HT}    & 86.19 & 84.92 & 12.49 & 11.46 \\
  \textbf{DS + TR}    & 86.64 & 85.29 & 13.44 & 11.47 \\
  \midrule
  \textbf{KI + WV}    & 73.79 & 72.63 & 16.58 & 14.03 \\
  \textbf{KI + W2}    & 74.56 & 73.32 & 15.40 & 13.68 \\
  \textbf{KI + WR}    & 74.38 & 73.11 & 15.88 & 13.61 \\
  \textbf{KI + XV}    & 74.12 & 72.96 & 14.52 & 11.67 \\
  \textbf{KI + HT}    & 73.76 & 72.54 & 16.65 & 13.26 \\
  \textbf{KI + TR}    & 83.05 & 82.58 & 14.55 & 11.68 \\
  \midrule
  \textbf{TA + WV}    & 76.94 & 75.58 & 15.78 & 12.29 \\
  \textbf{TA + W2}    & 76.31 & 74.92 & 16.66 & 12.93 \\
  \textbf{TA + WR}    & 77.26 & 75.88 & 15.26 & 12.54 \\
  \textbf{TA + XV}    & 76.64 & 75.28 & 14.79 & 11.43 \\
  \textbf{TA + HT}    & 77.04 & 75.84 & 15.13 & 11.61 \\
  \textbf{TA + TR}    & 74.06 & 72.89 & 14.13 & 12.92 \\
  \bottomrule
\end{tabular}
\caption{Simple concatenation performance on non-speech testing videos for all VFM+SFM combinations. All values are reported in percentage (\%) and averaged over 5-fold cross-validation.}
\label{tab:5}
\end{table}

\begin{table}[!h]
\setlength{\tabcolsep}{8pt}
\centering
\scriptsize
\begin{tabular}{l|cc|cc}
  \toprule
  \textbf{Combinations} & \multicolumn{4}{c}{\textbf{Non‐speech Testing Videos}} \\
  \cmidrule(lr){2-5}
  & \textbf{A \(\uparrow\)} & \textbf{F1 \(\uparrow\)} & \textbf{R \(\downarrow\)} & \textbf{M \(\downarrow\)} \\
  \midrule
  \multicolumn{5}{c}{\textbf{VFM + SFM}} \\
  \midrule
  \textbf{Vi + WV}   & 86.46 & 85.59 &  7.01 & 5.88 \\
  \textbf{Vi + W2}   & 85.94 & 85.11 &  7.33 & 5.50 \\
  \textbf{Vi + WR}   & 87.12 & 86.25 &  6.91 & 5.79 \\
  \textbf{Vi + XV}   & 86.29 & 85.44 &  6.70 & 5.64 \\
  \textbf{Vi + HT}   & 87.05 & 86.18 &  7.11 & 5.58 \\
  \textbf{Vi + TR}   & 85.78 & 84.95 &  6.35 & 4.72 \\
  \midrule
  \textbf{V2 + WV}   & 88.89 & 88.04 &  7.26 & 5.25 \\
  \textbf{V2 + W2}   & 88.21 & 87.93 &  7.26 & 5.32 \\
  \textbf{V2 + WR}   & 89.26 & 88.38 &  6.85 & 5.09 \\
  \textbf{V2 + XV}   & 88.74 & 87.89 &  \cellcolor{blue!25}\textbf{5.51} & 4.12 \\
  \textbf{V2 + HT}   & 89.55 & 88.64 &  7.10 & 5.07 \\
  \textbf{V2 + TR}   & 89.12 & 88.24 &  5.63 & \cellcolor{green!25}\textbf{3.91} \\
  \midrule
  \textbf{VV + WV}& 86.46 & 85.59 &  9.37 & 7.55 \\
  \textbf{VV + W2}& 85.94 & 85.11 &  9.19 & 7.09 \\
  \textbf{VV + WR}& 87.12 & 86.25 &  9.41 & 7.61 \\
  \textbf{VV + XV}& 86.29 & 85.44 &  9.26 & 6.32 \\
  \textbf{VV + HT}& 87.05 & 86.18 &  9.96 & 7.27 \\
  \textbf{VV + TR}& 85.78 & 84.95 &  8.97 & 7.19 \\
  \midrule
  \textbf{DS + WV}   & 91.89 & 91.02 &  6.80 & 4.91 \\
  \textbf{DS + W2}   & 91.63 & 90.77 &  6.70 & 4.72 \\
  \textbf{DS + WR}   & \cellcolor{yellow!25}\textbf{92.18} & \cellcolor{yellow!25}\textbf{91.23} &  6.34 & 4.48 \\
  \textbf{DS + XV}   & \cellcolor{green!25}\textbf{92.41} & \cellcolor{green!25}\textbf{91.47} &  \cellcolor{yellow!25}\textbf{5.80} & \cellcolor{yellow!25}\textbf{4.06} \\
  \textbf{DS + HT}   & 91.76 & 90.91 &  6.53 & 5.13 \\
  \textbf{DS + TR}   & \cellcolor{blue!25}\textbf{92.58} & \cellcolor{blue!25}\textbf{91.63} &  \cellcolor{green!25}\textbf{5.55} & \cellcolor{blue!25}\textbf{3.84} \\
  \midrule
  \textbf{KI + WV}   & 77.81 & 76.98 &  8.92 & 7.16 \\
  \textbf{KI + W2}   & 78.63 & 77.79 &  8.84 & 7.01 \\
  \textbf{KI + WR}   & 78.27 & 77.41 &  8.69 & 6.99 \\
  \textbf{KI + XV}   & 78.45 & 77.66 &  7.03 & 6.04 \\
  \textbf{KI + HT}   & 77.92 & 77.09 &  8.67 & 6.98 \\
  \textbf{KI + TR}   & 88.15 & 87.08 &  7.53 & 6.01 \\
  \midrule
  \textbf{TA + WV}   & 80.88 & 80.03 &  8.27 & 6.56 \\
  \textbf{TA + W2}   & 81.66 & 80.81 &  7.92 & 6.29 \\
  \textbf{TA + WR}   & 81.44 & 80.59 &  8.64 & 6.63 \\
  \textbf{TA + XV}   & 81.39 & 80.52 &  7.52 & 5.90 \\
  \textbf{TA + HT}   & 80.97 & 80.12 &  8.35 & 6.32 \\
  \textbf{TA + TR}   & 78.52 & 77.65 &  7.54 & 5.83 \\
  \bottomrule
\end{tabular}
\caption{Performance on combinations of the proposed \textbf{DIVINE} framework across non-speech testing videos for VFM+SFM. All values are reported in percentage (\%) and averaged over 5-fold cross-validation.}
\label{tab:6}
\end{table}

\subsection{Visualization Analysis}
\label{VA}

\subsubsection{Confusion Matrices}
\label{CM}

\begin{figure*}[!ht]
    \centering
    \subfloat[]{%
        \includegraphics[width=0.3\textwidth]{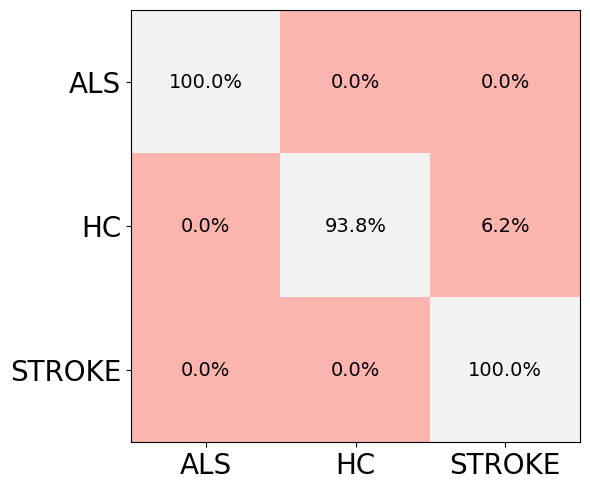}
    }
    \hspace{0.03\textwidth}
    \subfloat[]{%
        \includegraphics[width=0.3\textwidth]{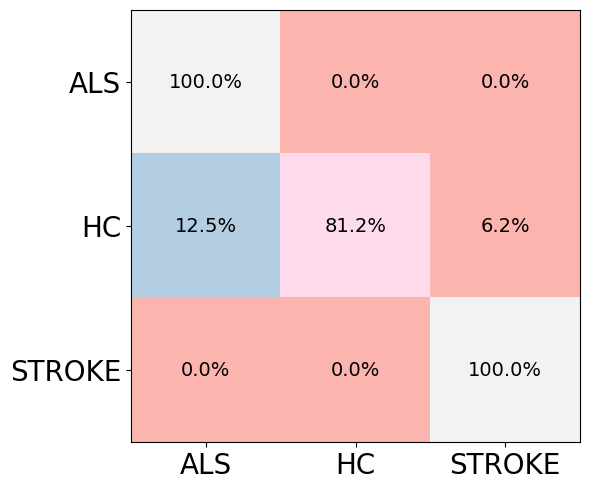}
    }\\[1ex]

    \subfloat[]{%
        \includegraphics[width=0.3\textwidth]{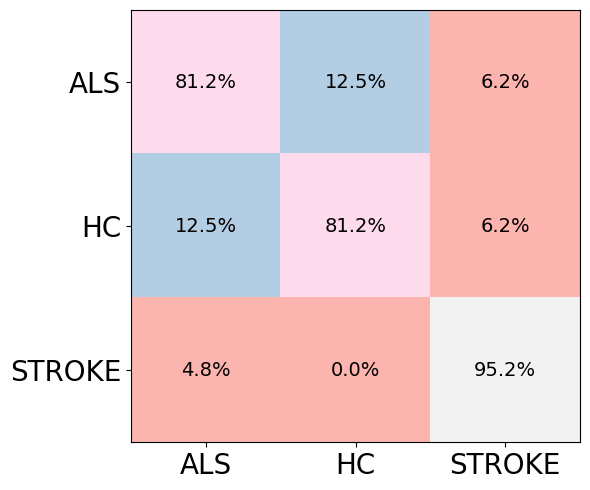}
    }
    \hspace{0.03\textwidth}
    \subfloat[]{%
        \includegraphics[width=0.3\textwidth]{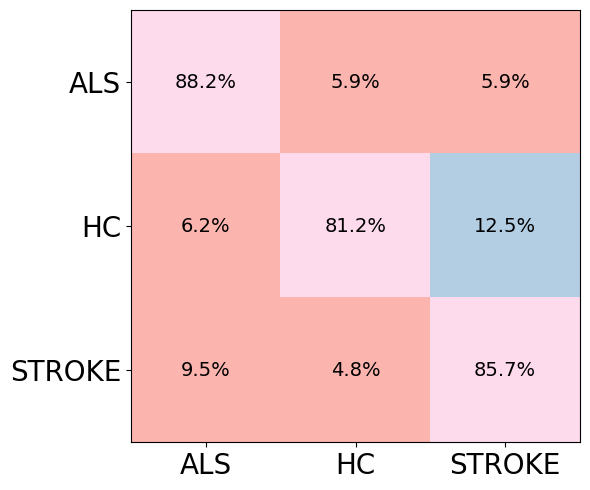}
    }\\[1ex]

    \subfloat[]{%
        \includegraphics[width=0.3\textwidth]{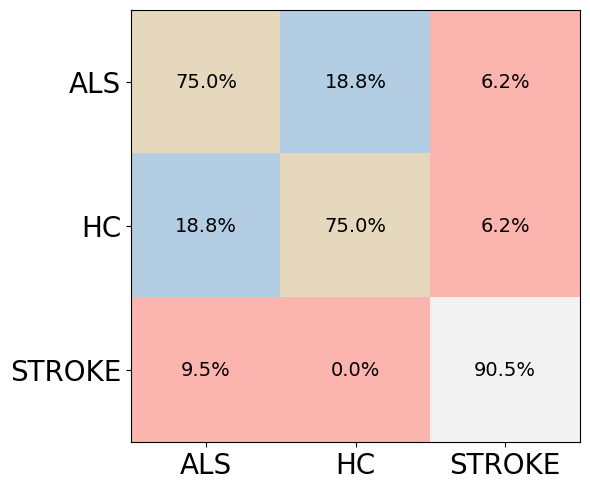}
    }
    \hspace{0.03\textwidth}
    \subfloat[]{%
        \includegraphics[width=0.3\textwidth]{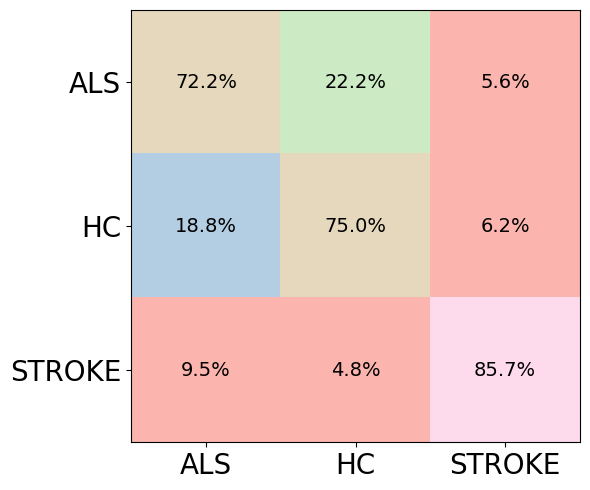}
    }\\[1ex]

    \subfloat[]{%
        \includegraphics[width=0.3\textwidth]{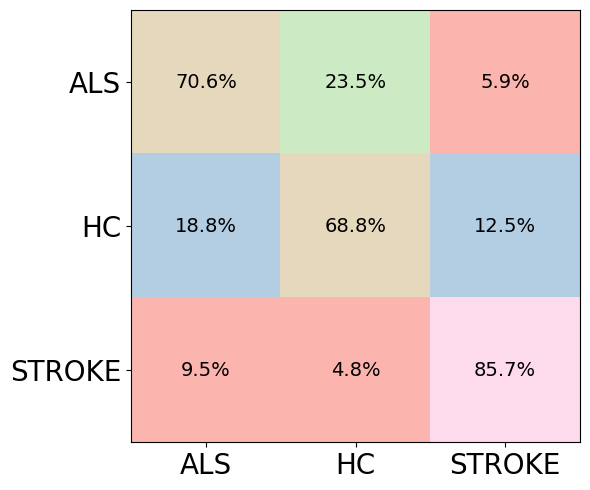}
    }
    \hspace{0.03\textwidth}
    \subfloat[]{%
        \includegraphics[width=0.3\textwidth]{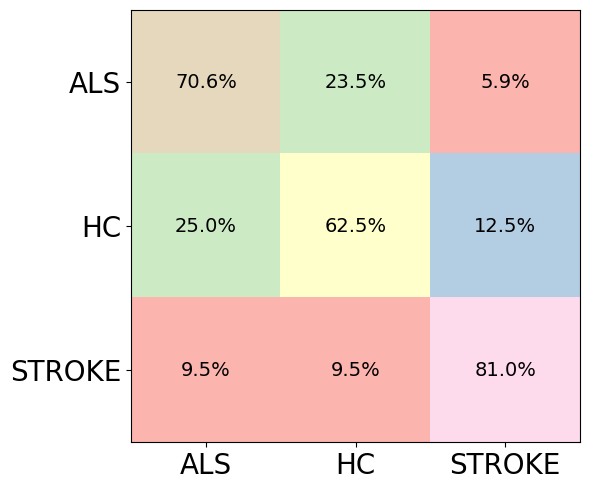}
    }
    \caption{Representing \textbf{DIVINE} configurations. Each displays true versus predicted class distributions across the combined diagnosis and severity categories: (a) DeepSeek‐VL2+TRILLsson; (b) DeepSeek‐VL2+X-vector; (c) DeepSeek‐VL2+X-vector (testing only video); (d) DeepSeek‐VL2+TRILLsson (testing only audio); (e) ViViT (Multitask); (f) WavLM; (g) Kinematic (Multitask); and (h) Kinematic (Classification). These matrices highlight classification consistency and error patterns for each fusion pairing.}
    \label{fig:cm_8}
\end{figure*}

\noindent Figure~\ref{fig:cm_8} presents confusion matrices for eight representative DIVINE configurations evaluated across our TNF test scenarios: (a) DeepSeek‐VL2 + TRILLsson; (b) DeepSeek‐VL2 + X-vector; (c) VideoMAE-V2 + TRILLsson; (d) VideoMAE-V2 + X-vector; (e) ViViT + TRILLsson; (f) ViViT + X-vector; (g) DeepSeek-VL2 + Wav2Vec 2.0; and (h) DeepSeek-VL2 + WavLM. These matrices illustrate classification consistency and error patterns across our key model pairings.  

\end{document}